\documentstyle[psfig,12pt]{article}
\def\TL{\hfil$\displaystyle{##}$}
\def\TR{$\displaystyle{{}##}$\hfil}
\def\TC{\hfil$\displaystyle{##}$\hfil}
\def\TT{\hbox{##}}
\def\seqalign#1#2{\vcenter{\openup1\jot
  \halign{\strut #1\cr #2 \cr}}}

\def\comment#1{}
\def\fixit#1{}

\def\tf#1#2{{\textstyle{#1 \over #2}}}

\def\mop#1{\mathop{\rm #1}\nolimits}

\def\coth{\mop{coth}}

\def\diag{\mop{diag}}

\def\overleftrightarrow#1{\vbox{\ialign{##\crcr
     $\leftrightarrow$\crcr\noalign{\kern-0pt\nointerlineskip}
     $\hfil\displaystyle{#1}\hfil$\crcr}}}

\def\lsim{\mathrel{\mathstrut\smash{\ooalign{\raise2.5pt\hbox{$<$}\cr\lower2.5pt\hbox{$\sim$}}}}}
\def\gsim{\mathrel{\mathstrut\smash{\ooalign{\raise2.5pt\hbox{$>$}\cr\lower2.5pt\hbox{$\sim$}}}}}

\def\sqr#1#2{{\vcenter{\vbox{\hrule height.#2pt
         \hbox{\vrule width.#2pt height#1pt \kern#1pt
            \vrule width.#2pt}
         \hrule height.#2pt}}}}
\def\square{\mathop{\mathchoice\sqr56\sqr56\sqr{3.75}4\sqr34\,}\nolimits}

\def\href#1#2{#2}  

\def\lbldef#1#2{\expandafter\gdef\csname #1\endcsname {#2}}
\def\eqn#1#2{\lbldef{#1}{(\ref{#1})}%
\begin{equation} #2 \label{#1} \end{equation}}
\def\eqalign#1{\vcenter{\openup1\jot
    \halign{\strut\span\TL & \span\TR\cr #1 \cr
   }}}

\def\comment#1{  \begin{raggedright}{\tt [#1]}\end{raggedright}}
\begin{document}
\baselineskip=14.5pt
\pagestyle{plain}
\setcounter{page}{1}
\renewcommand{\thefootnote}{\fnsymbol{footnote}}

\begin{titlepage}

\begin{flushright}
HUTP-99/A048 \\
MIT-CTP-2903 \\
hep-th/9909134
\end{flushright}
\vfil

\begin{center}
{\huge Modeling the fifth dimension}\\[8pt]
{\huge with scalars and gravity}
\end{center}

\vfil
\begin{center}
{\large O. DeWolfe,$^{1*}$ D.Z. Freedman,$^{2*}$ S.S. Gubser,$^{3*}$ and
A. Karch$^{1*}$}
\end{center}

$$\seqalign{\span\TL & \span\TT}{
^1 & Center for Theoretical Physics,
  \cr\noalign{\vskip-1.5\jot}
   & Massachusetts Institute of Technology, Cambridge, MA  02139-4307, USA 
  \cr\noalign{\vskip0.5\jot}
^2 & Department of Mathematics and Center for Theoretical Physics,
  \cr\noalign{\vskip-1.5\jot}
   & Massachusetts Institute of Technology, Cambridge, MA  02139-4307, USA 
  \cr\noalign{\vskip0.5\jot}
^3 & Lyman Laboratory of Physics, Harvard University, Cambridge, MA
02138, USA 
  \cr\noalign{\vskip0.5\jot}
}$$
\vfil

\begin{center}
{\large Abstract}
\end{center}

\noindent 
 A method for obtaining solutions to the classical equations for scalars
plus gravity in five dimensions is applied to some recent suggestions for
brane-world phenomenology.  The method involves only first order
differential equations.  It is inspired by gauged supergravity but does not
require supersymmetry.  Our first application is a full non-linear
treatment of a recently studied stabilization mechanism for inter-brane
spacing.  The spacing is uniquely determined after conventional fine-tuning
to achieve zero four-dimensional cosmological constant.  If the fine-tuning
is imperfect, there are solutions in which the four-dimensional branes are
de Sitter or anti-de Sitter spacetimes.  Our second application is a
construction of smooth domain wall solutions which in a well-defined limit
approach any desired array of sharply localized positive-tension branes.
As an offshoot of the analysis we suggest a construction of a supergravity
c-function for non-supersymmetric four-dimensional renormalization group
flows.

The equations for fluctuations about an arbitrary scalar-gravity
background are also studied.  It is shown that all models in which the
fifth dimension is effectively compactified contain a massless
graviton.  The graviton is the constant mode in the fifth dimension.
The separated wave equation can be recast into the form of
supersymmetric quantum mechanics.  The graviton wave-function is then
the supersymmetric ground state, and there are no tachyons.

\vfil
\begin{flushleft}
September 1999
\end{flushleft}

{\vskip 5pt \footnoterule\noindent
{\footnotesize $\,^*$\ {\tt odewolfe@ctp.mit.edu, dzf@math.mit.edu,
ssgubser@born.harvard.edu, karch@mit.edu}}}
\end{titlepage}
\newpage
\baselineskip=15.5pt

\section{Introduction}

Phenomenologists have recently studied higher dimensional
gravitational models containing one or more flat 3-branes embedded
discontinuously in the ambient geometry.  Scenarios with two 3-branes
provide an explanation of the large hierarchy between the scales of
weak and gravitational forces and contain a massless $2^{++}$ mode
which reproduces Newtonian gravity at long range on the branes
\cite{RS1,RS2}. In the following paper we present results of our study
of models of this type: specifically, results on the smoothing of
discontinuities and stabilization of inter-brane spacings in
5-dimensional models with gravity and a scalar field.  The issue of
fine-tuning in such models is also addressed. We also discuss the
fluctuation equations in these models somewhat differently from
treatments in the recent literature.

The centerpiece of this work is a supergravity-inspired approach to
obtain exact solutions of the nonlinear classical field equations in
gravity-scalar-brane models which is valid even without
supersymmetry. After a brief introduction to the technical issues in
section~\ref{Issues}, this approach is presented in section~\ref{Wise}
and applied to a class of models containing one positive and one
negative tension brane \cite{RS1} with compact $S_1/{\bf Z}_2$
geometry in the fifth dimension.  Stabilization of the brane spacing
is a generic feature of these models, but it is not guaranteed that
the branes will be flat.  Indeed, obtaining flat branes requires a
fine-tuning of the model precisely equivalent to setting the
four-dimensional cosmological constant to zero, and if the fine-tuning
is imperfect, the induced metric on the branes will be de Sitter space
or anti-de Sitter space.  The stabilization mechanism is a
generalization of the work of \cite{wise1}; however, our treatment
also includes back-reaction of the classical scalar profile.  An explicit
model is presented in section~\ref{Example}.

In section~\ref{Smooth} we obtain smooth solutions of gravity-scalar models
which approach discontinuous brane geometries in a certain ``stiff limit.''
Any array containing only positive tension branes can be smoothed in this
way.  We also remark on the usefulness of our first-order formalism for the
description of supergravity duals to renormalization group flows.

Our constructions have some parallels in earlier supergravity domain wall
literature (see \cite{CveticSoleng} for a review).  There are also
similarities with more recent literature, for example
\cite{Ovrut,Kehagias}.

In section~\ref{Fluctuate} we discuss the equations for linear
fluctuations about a gravity-scalar-brane configuration. We use the
axial gauge and a parameterization in which the 4-dimensional graviton
appears universally as a constant mode in the fifth dimension. This
mode is normalizable since that dimension is either manifestly or
effectively compact. The graviton equation can be transformed into the
form of a Schr\"odinger equation in supersymmetric quantum
mechanics. The graviton is the supersymmetric ground state, so there
is no lower energy state which would be a tachyon in the present
context.

\section{The issues}
\label{Issues}

We start with the five-dimensional gravitational action
\eqn{SGRAV}{S = \int d^5x \, \sqrt{g} \left[ -{1 \over 4} R + 
   {3 \over L^2} \right] \,,}
in $+----$ signature.  The most general
five-dimensional metric with four-dimensional Poincar\'e symmetry is
\eqn{MetricAnsatz}{ds^2 = e^{2A(r)} \, \eta_{ij} \, dx^i dx^j - dr^2
\,,}
with $\eta_{ij}=\diag\{1,-1,-1,-1,-1\}$.  Anti-de Sitter space is the
solution of the field equations of \SGRAV\ with $A(r) = -r/L$. This
metric describes a Poincar\'e coordinate patch in $AdS_5$ with
boundary region $r \rightarrow -\infty$ and Killing horizon region
$r\rightarrow +\infty$.

The basic positive tension brane considered in \cite{RS1,RS2} is given
by $A(r)= -|r|/L$. This can be thought of as the discontinuous (in
first derivative) pasting of the horizon halves of two Poincar\'e
patches with the 3-brane at $r=0$. One can obtain this as the solution
of the field equations for an action consisting of \SGRAV\ plus a
brane tension term:
  \eqn{SBRANE}{ S_{brane} = -\sum_{\alpha} \int d^4x \, dr \,
   \sqrt{|\det g_{ij}|} \, \lambda_{\alpha} \, \delta(r - r_{\alpha})
   \,. }
Here we have generalized to any number of branes;
$g_{ij}$ is the metric induced on each brane by the ambient
metric $g_{\mu\nu}$.  For a single brane at 
$r_1=0$ with brane tension $\lambda_1$, the $AdS$ scale must be
related by $1/L = \lambda_1/3$ to achieve a solution in which the
induced metric is flat.  This constraint represents a fine-tuning
which is precisely equivalent to setting the four-dimensional
cosmological constant equal to zero.

\begin{figure}
   \centerline{\psfig{figure=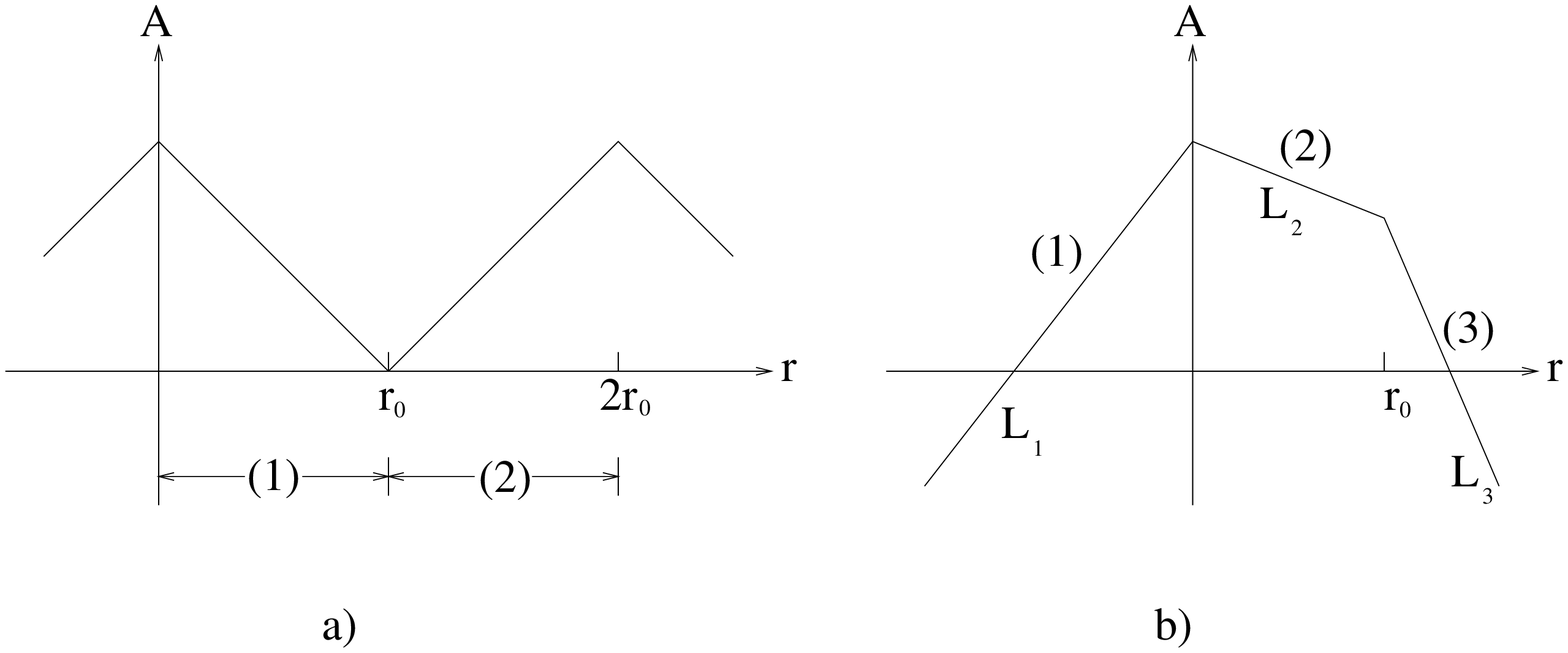,width=5in}}
    \caption{a) $A$ as a function of $r$ for the $S^1/{\bf Z}_2$ geometry,
with one positive and one negative tension brane, each at a fixed point of
${\bf Z}_2$.  b) $A$ as a function of $r$ for two positive tension branes
in an infinite fifth dimension.}\label{figA}
  \end{figure}

One can obtain a system of one positive and one negative tension brane
\cite{RS1} by considering two branes in \SBRANE\ with $\lambda_2=-\lambda_1$
and $r_2=r_0$.  This leads to the piece-wise linear scale function $A(r)$
shown in figure~\ref{figA}a. The fifth dimension is then periodic with
period $2r_0$ and there is a reflection symmetry under $r\rightarrow -r$.
This is the $S^1/{\bf Z}_2$ situation originally considered in
\cite{PolchinskiWitten,Horava}.

Another possibility is to consider \cite{Lykken} a second positive
tension brane, which admits a solution for $A(r)$ shown in
figure~\ref{figA}b. In this case, the bulk action \SGRAV\ must be
changed to admit different scales $L_1$, $L_2$, $L_3$ in the three
spatial regions. The scales are related to the brane tensions by
$1/L_1 +1/L_2 = 2\lambda_1/3$ and $1/L_3-1/L_2 = 2\lambda_2/3
>0$.  Again these relations must be regarded as fine-tunings absent a
dynamical mechanism by which they arise.

There are solutions of the equations of motion for any choice of the
inter-brane spacing $r_0$ in both scenarios above, so it is important
to ask whether there is any principle which fixes or stabilizes the
value of $r_0$. A first thought is that the total action integral of
the configuration might depend on $r_0$, reflecting an imbalance of
forces on the two 3-branes, and therefore could be minimized. However
it will be shown in the next section that the action vanishes for all
$r_0$, which apparently reflects the fact that the ``output'' value of
the classical four-dimensional cosmological constant vanishes, as is
consistent with the ``input'' value assumed when we considered
solutions containing flat 3-branes. In later sections we discuss
models in which a real scalar field $\phi$ with potential $V(\phi)$ is
coupled to gravity with brane tensions $\lambda_{\alpha}(\phi)$
depending on $\phi$. For a given choice of $V(\phi)$ and
$\lambda_{\alpha}(\phi)$, it is generally the case that the brane
spacing $r_0$ is uniquely determined.

Discontinuous solutions of field equations would be less artificial if
they could be obtained as a limit of smooth configurations. In 
section~\ref{Smooth} we present coupled scalar-gravity models with potential
$V(\xi)$ (and no branes initially present).  In these models the
scalar $\xi$ plays a different role, that of an auxiliary field, and
hence is given a different symbol.  The models have smooth domain wall
solutions which approach any desired discontinuous configuration of
positive tension branes as a scale parameter in $V(\xi)$ is varied.
Other parameters in $V(\xi)$ determine the inter-brane spacing
(e.g.\ $r_0$) and $AdS$ scales (e.g. $L_i$) of the limiting solution,
and the solutions have zero total action at all stages of the limiting
procedure.  The scalar $\xi$ is effectively frozen in the ``stiff''
limit of discontinuous branes.

Only positive tension brane configurations can be smoothed in this
way. A negative tension brane effectively has negative energy which
cannot be modeled in a conventional gravitational theory. Nevertheless
a negative tension brane is consistent with micro-physical
requirements if it is located at the fixed point of a discrete group
action.  The crucial point is that transverse fluctuations are then
projected out; otherwise they would have negative kinetic terms.

\section{The Goldberger-Wise mechanism}
\label{Wise}

It was proposed in \cite{wise1} that the dynamics of a scalar field
could stabilize the size of an extra dimension in the brane-world
scenario of \cite{RS1}.  The mechanism was to have a scalar $\phi$
with some mass in the bulk of a five-dimensional spacetime and some
potentials $\lambda_1(\phi)$ and $\lambda_2(\phi)$ on two
four-dimensional branes at the boundaries of this spacetime.  Such a
situation might be realized in the context of type~I$'$ string theory
\cite{Dai, PolchinskiWitten}, the Horava-Witten version of the
heterotic string \cite{Horava}, or some more ornate string theory
realization of the basic scenario of \cite{RS1}: in all cases,
spacetime has the topology ${\bf R}^{3,1} \times S^1/{\bf Z}_2$.  The
claim of \cite{wise1} is that stabilization of the length of the
interval $S^1/{\bf Z}_2$ can be achieved without fine-tuning the
parameters of the model (namely the mass of the scalar and the
potentials $\lambda_1(\phi)$ and $\lambda_2(\phi)$).

The analysis presented in \cite{wise1} neglected back-reaction of the
scalar field on the metric as well as the effect of different scalar
VEV's on the tensions of the branes.  The aim of this section is to
include these effects exactly.  To achieve a static solution with
$3+1$-dimensional Poincar\'e invariance to the full
gravity-plus-scalar-plus-branes equations, one fine-tuning is
necessary.  This fine-tuning amounts to setting the four-dimensional
cosmological constant to zero.  

The fine-tuning is somewhat different from the ones discussed in
\cite{stein,shirman}.  In \cite{stein} it was argued for a theory with
only gravity in the bulk that a nonzero four-dimensional cosmological
constant must necessarily be accompanied by rolling moduli
(corresponding to changing brane separations).  In \cite{shirman} it
was conjectured that a state with nonzero cosmological constant might
relax to zero cosmological constant, again through evolution of some
moduli specifying a brane configuration: in short, it was suggested
that an appropriate brane dynamics might fine-tune itself to zero
cosmological constant.  We will find a more conventional alternative:
there is generically a solution which is a warped product of a
maximally symmetric four-dimensional spacetime and an interval.  The
four-dimensional spacetime can be flat Minkowski spacetime, de Sitter
spacetime, or anti-de Sitter spacetime, and which is chosen depends on
the details of the scalar potentials in the bulk and on the branes.
Roughly speaking, one can construct a four-dimensional effective
potential $V_{\rm eff}$ whose extremal value determines the
cosmological constant.  There is no obvious dynamical principle in the
absence of supersymmetry which seems capable of forcing $V_{\rm eff} =
0$.  In particular, the presence of a fifth dimension simply does not
constrain the extremal value of $V_{\rm eff}$.  From a certain
viewpoint this should not come as a surprise: brane-world scenarios
must reduce at low energies to a four-dimensional gravity-plus-matter
theory, including some brane moduli with some potential, and it would
seem rather accidental than otherwise for this potential to enjoy a
fantastic property like zero extrema.

\subsection{A solution generating technique}
\label{Generate}

We generalize the action \SGRAV\ + \SBRANE\ to include a scalar field
$\phi(x^i,r)$:
  \eqn{Sfive}{
   S = \int_M d^4 x dr \, \sqrt{|\det g_{\mu\nu}|} \left[ -{1 \over 4} R + 
    {1 \over 2} (\partial\phi)^2 - V(\phi) \right] - 
    \sum_\alpha \int_{B_\alpha} d^4 x \, 
      \sqrt{|\det g_{ij}|} \lambda_\alpha(\phi) \ ,
  }
 where $M$ is the full five-dimensional spacetime and $B_\alpha$ is
the codimension one hypersurface where each brane is located.  It will
always be assumed that the branes are at definite values of $r$, so
that the $x^i$ are perpendicular to the brane hypersurfaces.

The solution generating method described in this section could be
applied to a fairly general setup with many codimension one branes on
a finite or infinite interval.  In this section our focus will be the
case of a finite interval $S^1/{\bf Z}_2$ where the only branes are
the ones at the ends of the interval.  We will work in the
``upstairs'' picture: ${\bf Z}_2$-symmetric configurations on the
circle $S^1$.  The bulk integration will extend over the entire $S^1$.
Properly speaking, the action should be cut in half after this
integration.  This can be achieved simply by setting $G_5 = 1/8\pi$
rather than $1/4\pi$.

We will initially assume a five-dimensional metric of the form
\MetricAnsatz.  We also assume that the scalar depends only on $r$.  These
assumptions follow if one demands a solution with $3+1$-dimensional
Poincar\'e invariance.  We will later generalize slightly by replacing
$\eta_{ij}$ with a de Sitter or anti-de Sitter metric.  It is
straightforward to obtain the Ricci tensor:
  \eqn{ricciFive}{
   R_{ij} = e^{2A} \, (4 A'^2 + A'') \, \eta_{ij} \qquad 
   R_{55} = -4 A'^2 - 4 A'' \,,
  }
 and to show that the equations of motion are
  \eqn{EOMS}{\eqalign{
   \phi'' + 4 A' \phi' &= {\partial V(\phi) \over \partial\phi} +
    \sum_\alpha {\partial \lambda_\alpha(\phi) \over \partial \phi} 
      \delta(r-r_\alpha)  \,, \cr
   A'' &= -{2 \over 3} \phi'^2 - {2 \over 3} \sum_\alpha 
    \lambda_\alpha(\phi) \delta(r-r_\alpha)  \,, \cr
   A'^2 &= -{1 \over 3} V(\phi) + {1 \over 6} \phi'^2 \ .
  }}
 We generally use primes to denote $d/dr$.  The last of the equations in
\EOMS\ is the usual zero-energy condition that follows from diffeomorphism
invariance.  If one differentiates it with respect to $r$, the result can
be shown to vanish identically if the first two equations are satisfied.

By integrating the first two equations on a small interval
$(r_\alpha-\epsilon,r_\alpha+\epsilon)$ one can derive the jump
conditions
  \eqn{Jump}{
   A'\Big|^{r_\alpha+\epsilon}_{r_\alpha-\epsilon} 
     = -{2 \over 3} \lambda_\alpha(\phi(r_\alpha))  \,, \qquad
   \phi'\Big|^{r_\alpha+\epsilon}_{r_\alpha-\epsilon} 
     = {\partial \lambda_\alpha \over \partial\phi}(\phi(r_\alpha)) \ .
  }
 If these conditions are satisfied at each brane, and if the first and
third equations of \EOMS\ are satisfied away from the branes, then we
have a consistent solution of the equations of motion everywhere.

Unfortunately we are still left with a difficult non-linear set of
equations.  We have been able to take advantage of one integral of the
motion (namely the zero-energy condition) to eliminate $A''$, and if
we wished we could eliminate $A'$ algebraically in the $\phi$ equation
by using the zero-energy condition, but we would still have a
difficult second order equation for $\phi$ with no further obvious
conserved quantities.  The purpose of this section is to exhibit a
general method of reducing the system \EOMS\ to three decoupled first
order ordinary differential equations, two of which are separable.
The method is inspired by supersymmetry but can be carried out
independent of it.  We should remark at the outset that our method is
only simple in the case of a single scalar $\phi$: one of our
differential equations has $\phi$ as the independent variable, and if
there were several scalars it would become a difficult partial
differential equation.

Suppose $V(\phi)$ has the special form
  \eqn{VWForm}{
   V(\phi) = {1 \over 8} 
    \left( \partial W(\phi) \over \partial \phi \right)^2 - 
     {1 \over 3} W(\phi)^2 \,,
  }
 for some $W(\phi)$.  Then it is straightforward to verify that a
solution to 
  \eqn{Weoms}{
   \phi' = {1 \over 2} {\partial W(\phi) \over \partial\phi} \,, \qquad
   A' = -{1 \over 3} W(\phi) \,,
  }
 is also a solution to \EOMS, provided we have
  \eqn{WlambdaRel}{
   {1 \over 2} W(\phi)\Big|^{r_\alpha+\epsilon}_{r_\alpha-\epsilon} = 
    \lambda_\alpha(\phi(r_\alpha))  \,, \qquad
   {1 \over 2} {\partial W(\phi) \over \partial\phi}
    \Bigg|^{r_\alpha+\epsilon}_{r_\alpha-\epsilon} = 
   {\partial \lambda_\alpha \over \partial\phi}(\phi(r_\alpha)) \ .
  }
 (It was previously noted in \cite{brandhuber} that the jump
conditions could be satisfied in a specific model if the brane tension
was given {\it identically} by $W(\phi)$, which is a much stronger
constraint on the model than we assume.)  Potentials of the form \VWForm\
occur in five-dimensional gauged supergravity \cite{fgpw1}, and the
conditions \Weoms\ arise as conditions for unbroken supersymmetry: the
vanishing of the dilatino variation leads to the first equation in
\Weoms\ and gravitino variation leads to the second.

For us, the key observation is that, given $V(\phi)$, \VWForm\ can be
solved for $W(\phi)$, and there is one integration constant in the
solution.  Whether a gauged supergravity theory can be constructed so
that the supersymmetry conditions lead to any desired $W(\phi)$ is an
interesting question which we will not address in this paper.  (It
would also be amusing to ask whether one could come up with
interesting supersymmetry-breaking scenarios by starting with a
five-dimensional gauged supergravity and constructing a solution using
\Weoms\ with the ``wrong'' $W(\phi)$.)  The relevant point for the
analysis at hand is that \VWForm\ and \Weoms\ together have solutions
specified by three integration constants, one of which is the trivial
additive constant on $A$.  There are likewise three integration
constants for the solutions of \EOMS, and again one is the trivial
additive constant on $A$.  From this simple parameter count we may
expect that the space of solutions includes all possible solutions to
 \EOMS.\footnote{R.~Myers \cite{MyersNotes} has also noted that \VWForm\
and \Weoms\ can be used to generate kink solutions, independent of
supersymmetry.  In the study of RG flows in AdS/CFT he has considered an
example with cubic $W(\phi)$ which is similar to the single-brane solution
which we will discuss in section~\ref{Smooth}.}  Issues of global existence
and discrete ambiguities seem to be the only obstacles to realizing this
expectation.  These are best seen in a more definite framework, so we will
now proceed to our main example.

The rest of this section is devoted to the case where the only branes
are the ones at the ends of the interval $S^1/{\bf Z}_2$.  Again, we
work in the ``upstairs'' picture where these branes are realized as
kinks in $A(r)$ at the fixed points of ${\bf Z}_2$.  If the ${\bf
Z}_2$ reflection includes an orientifolding, then string theory allows
one of these two branes to have negative tension.  The negative
tension brane must be located at a fixed point of the discrete group
action: it does not introduce difficulties with negative kinetic terms
or unboundedness of energy because it is just part of a background,
not something which can be dynamically created anywhere in space.  We
fix the additive ambiguity on the variable $r$ by taking the positive
tension brane to be at $r=0$.  The negative tension brane then lives
at some $r_0$ (see figure~\ref{figA}) which is the modulus of the
theory that the mechanism of \cite{wise1} purports to stabilize.  The
physical parameters that go into the theory are the scalar potential
$V(\phi)$ and the tensions $\lambda_1(\phi)$ and $\lambda_2(\phi)$.
These are assumed to emerge from the microscopic physics (for instance
string theory) which leads to this five-dimensional picture in a
low-energy limit (that is, low-energy compared to string scale and
ten-dimensional Planck scale as well as any further compactification
scales).  A moduli stabilization mechanism would be regarded as
fine-tuned if one has to impose some relationship among $V(\phi)$,
$\lambda_1(\phi)$, and $\lambda_2(\phi)$ to achieve a static solution.

Before explaining how the solutions to \EOMS\ can be generated using
\VWForm\ and \Weoms, let us do a quick count of parameters and constraints
to show that a fine-tuning is necessary to obtain a static solution with
flat branes.  There are three integration constants for the $\phi$ equation
plus the zero-energy equation in \EOMS: they are $\phi(0)$, $\phi'(0)$, and
$A(0)$.  There is one additional parameter, namely $r_0$, so four
parameters in all.  There are four constraints coming {}from the two jump
conditions at the two branes.  Naively one would conclude that there is no
fine-tuning: four contraints on four parameters can generically be solved.
But $A(0)$ is completely irrelevant because $A(r)$ enters into the
equations of motion and the jump conditions only through its derivatives.
That leaves three parameters subject to four constraints: indeed
fine-tuned.  This fine-tuning is equivalent to the fine-tuning required in
a theory without scalars between the brane tensions and the bulk
cosmological constant.

We will now argue in detail that any solution of \EOMS\ can be written
as a solution to \VWForm\ and \Weoms\ with an appropriately chosen
$W(\phi)$.  It is necessary to choose $W$ odd under the ${\bf Z}_2$
symmetry, just because $A'$ is equal and opposite at the two points on
any given ${\bf Z}_2$ orbit away from the fixed points.  With this in
mind we can restrict our attention to region $a$ in figure~\ref{figA}.
The jump conditions become
  \eqn{JumpSpecial}{\seqalign{\span\TL & \span\TR \qquad & 
    \span\TL & \span\TR}{
   A'(\epsilon) &= -{1 \over 3} \lambda_1(\phi(0)) \,, &
   \phi'(\epsilon) &= {1 \over 2} 
    {\partial \lambda_1 \over \partial\phi}(\phi(0))  \,, \cr
   A'(r_0-\epsilon) &= {1 \over 3} \lambda_2(\phi(r_0))  \,,&
   \phi'(r_0-\epsilon) &= -{1 \over 2}
    {\partial \lambda_2 \over \partial\phi}(\phi(r_0)) \ .
  }}
 Plugging these relations into the zero energy condition, we learn
that
  \eqn{VlambdaRel}{\seqalign{\span\TL & \span\TR \quad & \span\TT}{
   {1 \over 8} \left( {\partial \lambda_1 \over \partial\phi} \right)^2 - 
    {1 \over 3} \lambda_1^2 &= V & at $\phi = \phi_1$  \,, \cr
   {1 \over 8} \left( {\partial \lambda_2 \over \partial\phi} \right)^2 - 
    {1 \over 3} \lambda_2^2 &= V & at $\phi = \phi_2$ \ ,
  }}
 where $\phi_1$ and $\phi_2$ are the values attained by $\phi(r)$ at
$r=0$ and $r=r_0$, respectively.  Notice these constraints have the
same form as \VWForm\ $\!,$ with the $\lambda_{\alpha}$ playing the
role of $W$.  For generic $V$ and
$\lambda_\alpha$, the equations \VlambdaRel\ admit only a discrete set
of solutions for $\phi_1$ and $\phi_2$.  Given the physical input into the
model, namely $V(\phi)$, $\lambda_1(\phi)$, and $\lambda_2(\phi)$, the
discrete values $\phi_1$, $\phi_2$ are the points in field space where flat
branes can be consistently inserted.

Let us now integrate the equation \VWForm\ and fix the single integration
constant by requiring $W(\phi_1) = \lambda_1(\phi_1)$.  Because of \VWForm\
we have ${\partial W \over \partial\phi}(\phi_1) = \pm {\partial \lambda_1
\over \partial\phi}(\phi_1)$, and the plus sign is guaranteed if we assume
that ${\partial W(\phi) \over \partial\phi}$ has the same sign as
${\partial \lambda_1 \over \partial\phi}(\phi_1)$ in the vicinity of $\phi
= \phi_1$.  The solution $(A'(r),\phi(r))$ of \Weoms\ subject to $\phi(0) =
\phi_1$ must coincide with the solution $(A'(r),\phi(r))$ of \EOMS\ subject
to $\phi(0) = \phi_1$ and $\phi'(0) = {1 \over 2} {\partial \lambda_1 \over
\partial\phi}(\phi_1)$, because both of them satisfy the same boundary
data.  This is enough to conclude that locally every solution of \EOMS\ can
be generated by solving \VWForm\ and \Weoms.  Global issues of the
existence and uniqueness of solutions to \VWForm\ and \Weoms\ are best
addressed with a specific model in hand.  We will return to these points in
section~\ref{Example}.

\begin{figure}
   \centerline{\psfig{figure=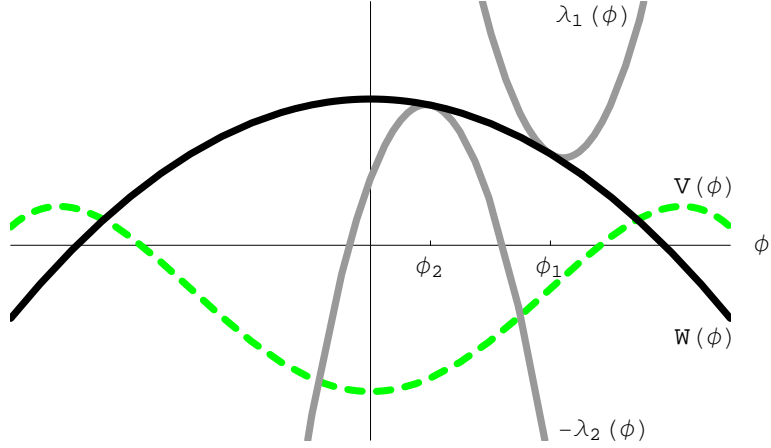,width=4in}}
   \caption{Sample $W$ (solid line), $V$ (dotted line), $\lambda_1$ and
$\lambda_2$ (grey lines) as functions of $\phi$.  By adjusting the
integration constant of \VWForm\ one can arrange for $\lambda_1$ to be
tangent to $W$, but then for $\lambda_2$ also to be tangent amounts to a
fine-tuning.}\label{figB}
  \end{figure}

Besides providing an efficient method for generating solutions to \EOMS,
the use of \VWForm\ and \Weoms\ also allows us to characterize in a simple
way how $\lambda_1(\phi)$, $\lambda_2(\phi)$, and $V(\phi)$ have to be
fine-tuned.  Having first fixed $W(\phi)$ in the manner described in the
previous paragraph, and then integrated \Weoms\ to obtain $\phi(r)$, we can
determine the position of the second brane by $\phi(r_0) = \phi_2$.  There
are no parameters left to fix (except for the trivial additive constant on
$A$), but we must still demand $W(\phi_2) = -\lambda_2(\phi_2)$ and
${\partial W \over \partial\phi}(\phi_2) = -{\partial \lambda_2 \over
\partial\phi}(\phi_2)$ in order that the jump conditions at the second
brane be satisfied.  Because of the defining property \VlambdaRel\ of
$\phi_2$, either one of these last two equations implies the other up to a
sign.  Thus there is precisely one fine-tuning, as expected from the
earlier parameter count.  The advantage of introducing $W$ is that the
fine-tuning condition can be expressed in terms of the solutions of the
single ordinary differential equation \VWForm\ (see figure~\ref{figB}).
 It should be kept in mind that we are working strictly at the
classical level.  If we tune parameters so that $W(\phi)$ and
$-\lambda_2(\phi)$ are tangent, then loop corrections to
$\lambda_{\alpha}(\phi)$ and $V(\phi)$ must be expected to spoil the
relation.  

It is true that if this fine-tuning can be achieved, there is no
cosmological constant allowed in the four-dimensional action.  A quick way
to see this is to show that the lagrangian is a total derivative with
respect to $r$ when \VWForm, \Weoms, and \WlambdaRel\ are satisfied: then
the four-dimensional lagrangian must vanish.\footnote{Since we assumed
$3+1$-dimensional Poincar\'e invariance in our ansatz from the start, zero
four-dimensional cosmological constant was guaranteed.  The following
computation is therefore only a consistency check.}  Let us define
  \eqn{hatW}{\eqalign{
   \hat{W}(\phi,r) = \left\{ 
    \seqalign{\span\TR \quad & \span\TT}{
     W(\phi) & for $0<r<r_0$  \cr
     -W(\phi) & for $r_0<r<2r_0$\ ,} \right. 
  }}
 which is appropriately ${\bf Z}_2$ odd.  Then it is straightforward
to show that
  \eqn{LagSimp}{\eqalign{
   {\cal L} &= \sqrt{|\det g_{\mu\nu}|} \left[ 
      -{1 \over 4} R + {1 \over 2} (\partial\phi)^2 - V(\phi) \right] -
      \sum_\alpha \sqrt{|\det g_{ij}|} \lambda_\alpha(\phi) 
       \delta(r-r_\alpha)  \cr
     &= e^{4 A} \left[ 3 \left( A' + {1 \over 3} \hat{W} \right)^2 - 
      {1 \over 2} \left(\phi' - 
       {1 \over 2} {\partial \hat{W} \over \partial\phi} \right)^2 + 
      {1 \over 2} {\partial \hat{W} \over \partial r} - 
      \sum_\alpha \lambda_\alpha \delta(r-r_\alpha) \right] \cr
     &\qquad - 
      {d \over dr} \left( e^{4 A} \left[ 2 A' + {1 \over 2} \hat{W} \right]
       \right) \ .
  }}
 In \LagSimp\ we have used \VWForm\ (with $W$ replaced by $\hat{W}$)
but not \Weoms.  If the perfect squares in \LagSimp\ vanish, then we
have
  \eqn{Whatlambda}{
   {1 \over 2} {\partial \hat{W} \over \partial r} = 
    W(\phi_1) \delta(r-r_1) - W(\phi_2) \delta(r-r_2) = 
    \sum_\alpha \lambda_\alpha \delta(r-r_\alpha) \,,
  }
 where in the second equality we have used the jump conditions,
\WlambdaRel.  In comparing with \hatW, recall that by convention $r_1=0$
and $r_2=r_0$.

The form of \LagSimp\ makes it clear that \Weoms\ are indeed a sort of
BPS condition for solutions of \EOMS.  However, because the perfect
squares in \LagSimp\ come in with opposite signs, there is no obvious
analog of a Bogomolnyi bound.  Another important implication of \LagSimp\
is that the total action of any configuration of flat branes vanishes.
This is even true of non-periodic arrays provided $A \to -\infty$ as $r \to
\pm\infty$.

\subsection{Non-zero cosmological constant}
\label{Nonzero}

The fine-tuning to achieve zero cosmological constant was already commented
on in \cite{wise1}.  The purpose of this section is show that if the
fine-tuning is imperfect, then there are solutions without rolling moduli
but where the metric on the branes is de Sitter space or anti-de Sitter
space.

Most of the analysis is similar to section~\ref{Generate}, so we will
be brief.  The metric ansatz is
  \eqn{deSitterAnsatz}{
   ds^2 = e^{2 A(r)} \bar{g}_{ij} dx^i dx^j - dr^2 \ ,
  }
 where $\bar{g}_{ij}$ is the metric of four-dimensional de Sitter
or anti-de Sitter spacetime: $\bar{R}_{ij} = -3 \bar{\Lambda}
\bar{g}_{ij}$, where $\bar{\Lambda}$ is the four-dimensional
cosmological constant (positive for de Sitter spacetime and negative for
anti-de Sitter spacetime).  Explicitly, we may write the four-dimensional
metrics as 
  \eqn{deSitterSpace}{\eqalign{
   dS_4 : & \quad \bar{g}_{ij} dx^i dx^j = dt^2 - 
    e^{2 \sqrt{\bar{\Lambda}} t} (dx_1^2 + dx_2^2 + dx_3^2)  \cr
   AdS_4 : & \quad \bar{g}_{ij} dx^i dx^j = e^{-2\sqrt{-\bar{\Lambda}} x_3} 
    (dt^2 - dx_1^2 - dx_2^2) - dx_3^2 \ .
  }}
 The five-dimensional Ricci tensor and the equations of motion are
  \eqn{ricciFiveCosm}{\seqalign{\span\TC}{
   R_{ij} = e^{2A} \left( 4 A'^2 + A'' - 3 \bar{\Lambda} e^{-2 A} 
    \right) \bar{g}_{ij} \,, \qquad 
   R_{55} = -4 A'^2 - 4 A''  \,, \cr
  \eqalign{
   \phi'' + 4 A' \phi' &= {\partial V(\phi) \over \partial \phi} + 
    \sum_\alpha {\partial \lambda_\alpha(\phi) \over \partial \phi} 
      \delta(r-r_\alpha)  \,,  \cr
   A'' + \bar{\Lambda} e^{-2A} &= 
    -{2 \over 3} \phi'^2 - {2 \over 3} \sum_\alpha 
    \lambda_\alpha(\phi) \delta(r-r_\alpha) \,, \cr
   A'^2 - \bar{\Lambda} e^{-2A} &= -{1 \over 3} V(\phi) + 
    {1 \over 6} \phi'^2 \ .
  }}}
 The jump conditions \Jump\ are unchanged.  Neither $A(r)$ nor
$|\bar{\Lambda}|$ can be determined unambiguously from the equations of
motion because they enter only in the combination $A(r) - {1 \over 2}
\ln |\bar{\Lambda}|$.  We will see that this combination is what
determines the four-dimensional cosmological constant in
four-dimensional Planck units.  We could adjust the additive constant
on $A$, if we so desired, to set $\bar{\Lambda} = 1$ for de Sitter
spacetime or $\bar{\Lambda} = -1$ for anti-de Sitter spactimee.  The
important point is not to count the magnitude of $\bar{\Lambda}$ as an
adjustable parameter separate from the additive constant on $A$.

Already from \ricciFiveCosm\ we can see why there should be a solution
with no fine-tuning of parameters.  The $\phi$ equation and the
zero-energy condition together have three integration constants, and
there is also the brane separation $r_0$.  Because $A$ itself rather
than just its derivatives enters into the equations \ricciFiveCosm,
the additive constant on $A$ is no longer trivial.  As before there
are four boundary conditions (two jump conditions at each brane), so
generically one expects a (locally) unique solution for any given
$V(\phi)$, $\lambda_1(\phi)$ and $\lambda_2(\phi)$.

The solution in the bulk (more precisely, in region $A$ of
figure~\ref{figA}) can still be obtained as a solution of a slightly
modified system of first order equations,\footnote{We are grateful to
Martin Gremm and Lisa Randall for pointing out to us an error in an earlier
version concerning these first order equations.}
\begin{eqnarray}
 \label{WCosmEOMS}
   A' &=& - \frac{1}{3} \, W \, \gamma(r) \,, \\
\nonumber
   \phi' &=& {1 \over 2 \gamma(r)} \, {\partial W \over \partial \phi} \,, \\
\nonumber
   V &=& \frac{1}{8 \gamma(r)^2} \left ( \frac{\partial W}{\partial \phi}
   \right )^2 - \frac{1}{3} \, W^2 \,,
\end{eqnarray}
which differ from \Weoms\ just by inclusion of the
factor \eqn{Fudgefactor}
{\gamma(r) \equiv \sqrt{1+ \frac{9 \bar{\Lambda}}{W(r)^2} \, e^{- 2 A(r)}}\,.} 
Note that this completely changes the character of the problem.  In
the case of zero cosmological constant, the first order equations
(\ref{VWForm}), (\ref{Weoms}) allowed us to find solutions for given $V$ directly by first integrating (\ref{VWForm}) to
solve for $W(\phi)$, then using the equations (\ref{Weoms}) to solve
consecutively for $\phi(r)$ and $A(r)$.  We now see that if  we do not
fine-tune the cosmological constant to zero, we obtain a complicated
non-linear first order system of differential equations for 3
functions $W(r)$, $\phi(r)$ and $A(r)$, now viewed as functions of a
single independent variable $r$, which we cannot simply solve for in
sequence.  $V(\phi)$ is still to be considered the information that is
put in from the Lagrangian, but its relationship with $W$ can no longer be isolated from the rest of the system.  Derivatives with respect to $\phi$ should
now be thought of as
\eqn{Phiderivative}
{\frac{\partial }{\partial \phi} = \frac{1}{\phi'(r)} \frac{\partial}
{\partial r}.}
To make this point more transparent, it is useful to rewrite
the system (\ref{WCosmEOMS}) as an autonomous system, that is
in the form
\eqn{Autonomous}{
A' = f(A, W,\phi), \, \qquad \phi' = g(A, W, \phi),
\, \qquad W' = h(A, W, \phi) \,,
}
where
\begin{eqnarray}
f(A,W,\phi) &=& -\frac{1}{3} \, \gamma(r) \, W(r) \,,\\ \nonumber
g(A,W,\phi) &=& \sqrt{2 \, V(\phi(r)) + \frac{2}{3} \, W(r)^2} \,, \\ 
\nonumber
h(A,W,\phi) &=& 2 \gamma(r) \left(2 \, V(\phi(r) + \frac{2}{3} \, W(r)^2 \right) \,.
\end{eqnarray}
While for a generic $V(\phi)$ this system will still be hard to solve,
it is very well suited for generating examples where $V(\phi)$ is
determined at the end.  For any given shape of the warp factor $A(r)$
one desires, one can find a potential that supports such a solution by
the following procedure: pick $A(r)$, calculate $A'(r)$, solve $A'=
-\frac{W}{3} \gamma$ algebraically for $W(r)$, and use $\phi' = \sqrt{
\frac{W'}{2 \gamma}}$ to obtain $\phi(r)$.  $V(r)$ can now simply be
determined by plugging in $W$ and $\phi$, and after inverting
$\phi(r)$ to $r(\phi)$ one obtains the desired $V(\phi)$. This
procedure for example can be used to generate fat branes (as we will
discuss them in later chapters for the
Minkowski case) with an AdS or dS worldvolume.
Note that this simple technique for generating examples is not possible
in the obvious first order system one could write down simply by
introducing one new variable $y$ with the one new defining
equation $y=\phi'$, as it is a standard technique for
converting a system of higher order equations into a first order system.

Assuming (\ref{WCosmEOMS}), the
jump conditions reduce to
  \eqn{JumpCosm}{\seqalign{\span\TL & \span\TR \qquad & \span\TL & \span\TR}{
   \lambda_1(\phi(0)) &= W(r=0) \gamma(0),  &
   {\partial \lambda_1 \over \partial\phi}(\phi(0)) &=
    \frac{1}{\gamma(0)} 
\frac{\partial W}{\partial \phi} \Bigg|_{\phi(0)}  \,, \cr
   \lambda_2(\phi(r_0)) &= W(r=r_0) \gamma(r_0),  &
   {\partial \lambda_2 \over \partial\phi}(\phi(r_0)) &=
    \frac{1}{\gamma(r_0)} 
\frac{\partial W}{\partial \phi} \Bigg|_{\phi(r_0)} \ .
  }}

If for a given $V(\phi)$ we fix $A(0)$ arbitrarily, then
the 5 other initial conditions, $\phi(0)$, $A'(0)$, $W(0)$,
$W'(0)$, $\phi(0)$, $\phi'(0)$  can be
determined up to discrete choices, using the 3 equations 
from (\ref{WCosmEOMS}) evaluated at $r=0$ and the 2 from the first line of
\JumpCosm.  Then \Autonomous\ can be solved unambiguously for
$\phi(r)$, $W(r)$ and $A(r)$.
$r_0$ is fixed by the last equality in
(\ref{WCosmEOMS}).  One is left with one condition, namely the third equality
in \JumpCosm.  It is a (very complicated) constraint on $A(0)$, which
generically will have only discretely many solutions.  The point is
that we wind up with exactly as many parameters as constraints, so it
doesn't take any fine-tuning to get a solution.

There does not seem to be a simple way to express the action as a sum
(or difference) of squares plus total derivatives, in analogy to
\LagSimp.  However it is straightforward to use the equations of
motion to show that 
  \eqn{LagSimpCosm}{\eqalign{
   {\cal L} &= \sqrt{|\det g_{\mu\nu}|} \left[ 
      -{1 \over 4} R + {1 \over 2} (\partial\phi)^2 - V(\phi) \right] -
      \sum_\alpha \sqrt{|\det g_{ij}|} \lambda_\alpha(\phi) 
       \delta(r-r_\alpha)  \cr
     &= \sqrt{|\det \bar{g}_{ij}|} 
         \left[ {3 \over 2} e^{2A} \bar{\Lambda} - 
         {d \over dr} \left( {1 \over 2} e^{4 A} A' \right) \right] \ .
  }}
 When ${\cal L}$ is integrated over the $S^1$ parametrized by $r$, it
must for consistency reduce to the four-dimensional lagrangian,
  \eqn{barLForm}{
   \bar{\cal L} = {1 \over 4 \pi G_4} \sqrt{|\det \bar{g}_{ij}|}
    \left[ -{1 \over 4} \bar{R} - {3 \over 2} \bar{\Lambda} \right] \ ,
  }
 evaluated on de Sitter or anti-de Sitter spacetime, where $\bar{R}_{ij} =
-3 \bar{\Lambda} \bar{g}_{ij}$, with $\bar{\Lambda}$ positive or negative,
respectively.  Comparison yields the relation
  \eqn{GLambdaRel}{
   {1 \over G_4} = 4\pi \int dr \, e^{2A} \ ,
  }
 where as usual the $r$ integration is over the whole of $S^1$.  For
consistency with observation we must demand the bound
  \eqn{CosmBound}{
    {1 \over G_4 |\bar{\Lambda}|} \gsim
    \left( {\ell_{\rm Hubble} \over \ell_{\rm 4d\ Planck}} \right)^2
    \approx 10^{120} \ .
  }
 In view of \GLambdaRel\ this translates to
  \eqn{BigTune}{
   4\pi \int dr \, e^{2 \left( A(r) - {1 \over 2} \ln|\bar{\Lambda}|
    \right)} \gsim 10^{120} \ .
  }
 The function $A(r) - {1 \over 2} \ln |\bar{\Lambda}|$ is fixed by
(\ref{WCosmEOMS}) and \JumpCosm\ once $V(\phi)$, $\lambda_1(\phi)$, and
$\lambda_2(\phi)$ are specified.  A dramatic fine-tuning in these
quantities is required to achieve \BigTune.

In general it is difficult to obtain solutions to
\ricciFiveCosm\ or (\ref{WCosmEOMS}) in closed form.  
We can however give a complete treatment of
the case where there is no scalar and $W$ is just a constant (namely the
square root of the bulk cosmological constant); see also
\cite{Nihei,Kaloper}.  In this case the only
equations we have to solve are the first equations in each line of
(\ref{WCosmEOMS}) and \JumpCosm.  The solutions can be expressed as follows:
  \eqn{deSitterSol}{\eqalign{
   & dS_4 : \quad \bar{\Lambda} > 0  \cr
   & \qquad e^A = \sqrt{\bar{\Lambda}} L \sinh {r_1-r \over L} \ , \quad
    \lambda_1 = {3 \over L} \coth {r_1 \over L} \ , \quad
    \lambda_2 = -{3 \over L} \coth {r_1-r_0 \over L} \cr
   & AdS_4 : \quad \bar{\Lambda} < 0  \cr
   & \qquad e^A = \sqrt{-\bar{\Lambda}} L \cosh {r_1-r \over L} \ , \quad
    \lambda_1 = {3 \over L} \tanh {r_1 \over L} \ , \quad
    \lambda_2 = -{3 \over L} \tanh {r_1-r_0 \over L}
  }}
 In the $dS_4$ case it is necessary to restrict $r_0 < r_1$.  The main
point which \deSitterSol\ demonstrates is the following.  Suppose one starts
with any fixed negative bulk cosmological constant, $-4/L^2$, and arbitrary
but specified $\lambda_1$ and $\lambda_2$, subject only to the constraint
that if one of the $\lambda_\alpha$ exceeds $3/L$ in magnitude, then the
other must also exceed $3/L$ in magnitude and be of the opposite sign.
Then there is a unique solution to \deSitterSol\ up to the usual ambiguity
between the additive constant on $A$ and the magnitude of $\bar{\Lambda}$.
Both $r_1$ and $r_0$ will be fixed in this solution, and so will the
combination $A - {1 \over 2} \ln |\bar{\Lambda}|$ which determines the
four-dimensional cosmological constant in Planck units.  The only exception
is when $\lambda_1 = -\lambda_2 = {3 \over L}$: in this case the branes are
flat, $r_1$ is a meaningless additive constant on $A$, and the brane
separation $r_0$ is {\it not} fixed.

The bulk solutions in \deSitterSol\ have vanishing Weyl tensor, hence they
are locally $AdS_5$.  All we have found, then, is an embedding of $AdS_4$
and $dS_4$ as codimension one hypersurfaces in $AdS_5$.  To verify this one
can find an explicit change of variables which brings the bulk metric into
the standard form
  \eqn{StandardTilde}{
   ds^2 = e^{-2\tilde{r}/L} (d\tilde{t}^2 - d\tilde{x}_1^2 - 
    d\tilde{x}_2^2 - d\tilde{x}_3^2) - d\tilde{r}^2 \ .
  }
 If we demand that the map from untilded to tilded coordinates be
orientation preserving, then the natural choice is 
  \eqn{VariableChange}{\seqalign{\span\TR & \quad \span\TR}{
   dS_4 : & \tilde{t} = -\sqrt{\bar{\Lambda}} \coth {r_1-r \over L} 
    e^{-\sqrt{\bar{\Lambda}} t} \ , \quad
   \tilde{r} = -\sqrt{\bar{\Lambda}} L t - L \log \sinh {r_1-r \over L} \ , \cr
    & \tilde{x}_1 = x_1 \ , \quad
      \tilde{x}_2 = x_2 \ , \quad
      \tilde{x}_3 = x_3  \cr
   AdS_4 : & \tilde{x}_3 = \sqrt{-\bar{\Lambda}} \tanh {r_1-r \over L} 
    e^{\sqrt{-\bar{\Lambda}} x_3} \ , \quad
    \tilde{r} = {\sqrt{-\bar{\Lambda}}} L x_3 - L \log \cosh {r_1-r \over L} 
      \ , \cr
    & \tilde{t} = t \ , \quad
      \tilde{x}_1 = x_1 \ , \quad
      \tilde{x}_2 = x_2 \ .
  }}
 Let us now focus on the $dS_4$ case with one positive and one negative
tension brane at the ends of the bulk.  A solution of the form
\deSitterSol\ maps to a strip of the $\tilde{t}-\tilde{r}$ plane between
two curves of the form $\tilde{t} = -c_\alpha e^{\tilde{r}/L}$.  Here $c_1$
and $c_2$ are positive constants.  Because $\partial/\partial\tilde{t}$ is
a Killing vector of the bulk geometry, we can trivially obtain a broader
class of solutions which have as their boundaries curves of the form
$\tilde{t} - \tilde{t}_\alpha = -c_\alpha e^{\tilde{r}/L}$, where now
$\tilde{t}_1$ and $\tilde{t}_2$ are additional constants, only one of which
can be set to $0$ through diffeomorphism freedom.  In these solutions the
proper distance between the branes is not constant.  In fact, generically
the branes intersect at some point, or they intersect the boundary of
$AdS_5$ at different points---or both.  In the latter case the graviton
bound state ceases to exist at some finite time as measured on the negative
tension brane.  This reinforces the intuition that brane-world cosmology
can encounter some curious pathologies.

The strategy of displacing one boundary by some distance along the flow of
a Killing vector of $AdS_5$ can also be applied to flat branes.  For
instance, one could shift the negative tension brane forward along the {\it
global} time of $AdS_5$ to obtain a new solution where the proper distance
between the branes is non-constant.  The positive and negative tension
branes would then intersect at some time in the distant past, and the
positive tension brane would again retreat to the true boundary of $AdS_5$
at a finite time as measured on the negative tension brane.  This is a
catastrophe since it means that gravity would cease altogether in four
dimensions: the four-dimensional Planck length would vanish.

\section{An explicit model}
\label{Example}

It is useful now to turn to an explicit example with non-trivial dynamics
for a single scalar.  For simplicity, we choose quadratic $W(\phi)$,
$\lambda_1(\phi)$, and $\lambda_2(\phi)$ which are tangent to one another
in the manner illustrated in figure~\ref{figB}.  Explicitly, 
  \eqn{ExplicitWVLambda}{\eqalign{
   W(\phi) &= {3 \over L} - b \phi^2  \,, \cr
   V(\phi) &= -{3 \over L^2} + \left( {b^2 \over 2} + {2b \over L} \right)
    \phi^2 - {b^2 \phi^4 \over 3}  \,, \cr
   \lambda_1(\phi) &= W(\phi_1) + W'(\phi_1) (\phi-\phi_1) + 
    \gamma_1 (\phi-\phi_1)^2 \,, \cr
   \lambda_2(\phi) &= -W(\phi_2) - W'(\phi_2) (\phi-\phi_2) +
    \gamma_2 (\phi-\phi_2)^2 \ .
  }}
 We stress that the physical properties of the model are summarized by
$V(\phi)$ and the $\lambda_\alpha(\phi)$: in the absence of supersymmetry,
there is no preferred choice of $W(\phi)$.  In section~\ref{Numerics} we
will analyze the different possible $W(\phi)$ that lead to the particular
quartic $V(\phi)$ exhibited in \ExplicitWVLambda.  Until then we will just
assume that the particular $W(\phi)$ that is tangent to $\lambda_1(\phi)$
happens to be the quadratic one shown in \ExplicitWVLambda.  We make this
assumption in order to obtain solutions in closed form.  The only physical
fine-tuning is the requirement that $-\lambda_2(\phi)$ is also tangent to
$W(\phi)$.  The quantities $L$, $b$, $\phi_1$, $\phi_2$, $\gamma_1$, and
$\gamma_2$ are parameters of the various potentials, and no dimensionless
ratio of them should be large if we want to preserve naturalness.

We will always assume that $\gamma_1$ and $\gamma_2$ are positive so that
the energetics of $\lambda_1$ and $\lambda_2$ tend to stabilize the
positions of the branes in field space.  We will usually assume $b > 0$ as
well.  It should be noted that $V(\phi)$ is unbounded below, as is common
and without pathology in $AdS$ supergravity.

\subsection{Analytical calculations}
\label{Analytic}

It is trivial to solve the first order equations \Weoms\ in the model
\ExplicitWVLambda\ to obtain
\eqn{ExplicitSoln}{\eqalign{
\phi(r) &= \phi_1\,  e^{-br} \,, \cr
A(r) &= a_0 - {r \over L} - {1 \over 6} \, \phi_1^2 \, e^{-2br} \ .
}}

The brane spacing is determined by the condition $br_0 =
\ln(\phi_1/\phi_2)$.  The difference $A(0)-A(r_0)$ gives the number of
$e$-foldings in discussions \cite{RS1,Lykken} of the gauge hierarchy
problem,\footnote{We assume that the four-dimensional and five-dimensional
Planck scales are comparable.  It is possible to relax this assumption
\cite{tjLi} since the additive constant on $A(r)$ is a free parameter.}
and one easily obtains
  \eqn{eFoldings}{A(0) - A(r_0) = {1 \over bL} \ln {\phi_1 \over \phi_2} - 
    {1 \over 6} (\phi_1^2-\phi_2^2) \ .
  }
 Phenomenologically one wants 
  \eqn{AWeakScale}{
   A(0)-A(r_0) \approx \ln {M_{\rm Planck} \over M_{\rm electroweak}} 
    \approx 37 \ .
  }
 If $b>0$, then $\phi_1^2-\phi_2^2 > 0$, and only first term can contribute
to the hierarchy.  This is conceivable if $bL$ is fairly small: for
instance, if $\phi_1/\phi_2 = e$ then one needs $bL \approx 1/37$.  If
$b<0$, then both terms in \eFoldings\ could contribute to the hierarchy.
One could for instance obtain an acceptable hierarchy by taking $bL = 1$,
$\phi_1 = 1$, and $\phi_2 = 15$.

The treatment of \cite{wise1} ignored back-reaction of the scalar profile
on the geometry.  Crudely speaking this means one should drop the second
term in \eFoldings\ since it came from a term proportional to the square of
the scalar field in \ExplicitSoln.  More precisely, (14) of \cite{wise1}
can be reproduced exactly by dropping the second term in \eFoldings\ and
identifying their $m^2 L^2$ with our $bL$ in the limit of small $bL$.  Thus
the analysis of \cite{wise1} was essentially adequate for the case $b > 0$,
where to obtain a large hierarchy one wants a bulk geometry which is not so
far from $AdS_5$ that the second term of \eFoldings\ is large.  However the
inclusion of back-reaction becomes quite important in the $b<0$ case, where
a large hierarchy can be most easily obtained via a geometry which deviates
strongly from $AdS_5$.

Any mechanism for generating large numbers must be probed for robustness.
We may ask, once the hierarchy \AWeakScale\ is obtained, how much can the
parameters change and still give the same weak scale to within errors?  For
definiteness, let us ask what change of parameters shifts $A(0)-A(r_0)$ by
no more than $0.02$: this would amount to a shift of the weak scale by two
percent, which is about the ratio of the $Z$ width to its mass.  In the
$b>0$ scenario we described above, a change of $\phi_1/\phi_2$ by about one
part in $2000$ changes the weak scale by two percent: multiplicative shifts
in this ratio are magnified by the factor $1/bL$.  In the $b<0$ scenario,
changing $\phi_2$ by about one percent changes the weak scale by two
percent.  Thus (superficially at least) the $b<0$ scenario is more robust.

\subsection{Numerics}
\label{Numerics}

We now change gears and refocus on \VWForm.  The purpose is to
illustrate the problem of selecting a superpotential $W(\phi)$ which
reproduces a given potential function $V(\phi)$. However, we shall be
content to explore this question only in the model of this section,
where $V(\phi)$ is given in \ExplicitWVLambda.  It is convenient to
rescale variables, partly to prepare for use of the MATLAB linked
program DFIELD5 \cite{Polking}.  We therefore define

\eqn{Rescale}{\eqalign{ \phi &= \sqrt{3 \over 8}\, t \,, \cr V(\phi) =
3 \, b^2 \, U(t) \,, \quad \quad U(t) &= -{1 \over b^2 L^2} + \left(
1+ {4 \over bL} \right) {t^2 \over 16} -{t^4 \over 64} \,, \cr W(\phi)
= 3 \, b \, x(t) \,, \quad \quad x_0(t) &= {1 \over bL} - {t^2 \over
8} \,.  }} We denote the rescaled preferred superpotential by $x_0(t)$
since we will consider other superpotentials corresponding to the
potential $U(t)$.

In this notation \VWForm\ takes the form \eqn{dXdTsquared}{ \left({dx
  \over dt}\right)^2 = x(t)^2 + U(t) \,. } There is a sign ambiguity
  in taking the square root which must be kept in mind, but we will
  discuss only the features of the differential equation which results
  from the positive root, namely \eqn{dXdT}{ {dx \over dt} = \sqrt{x^2
  - {1 \over b^2 L^2} + \left( 1+ {4 \over bL} \right) {t^2 \over 16}
  -{ t^4 \over 64}} \,.  } The equation is roughly like the energy
  equation in the mechanics problem of a particle in an inverted
  harmonic potential. As in mechanics there are forbidden regions of
  the $t-x$ plane where $x^2 +U(t)<0$. At a boundary of this region,
  which would be a turning point in a mechanics problem, the slope
  $\tf{dx}{dt}$ vanishes. According to the general theory of
  first order differential equations there is a unique solution curve
  through every point not in a forbidden region. The inequality
  \eqn{}{\left| {dx \over dt} \right| \leq |x| + \sqrt{|U(t)|} \,,}
  shows that no solution reaches $|x| =\infty$ at a finite field
  value.

The DFIELD5 program quite rapidly provides a reasonable global and
quantitative picture of the space of solutions. The quantities of our
problem depend only on the single dimensionless parameter $bL$, and we
set $bL=1$ in our numerical work.

  \begin{figure}
   \centerline{\psfig{figure=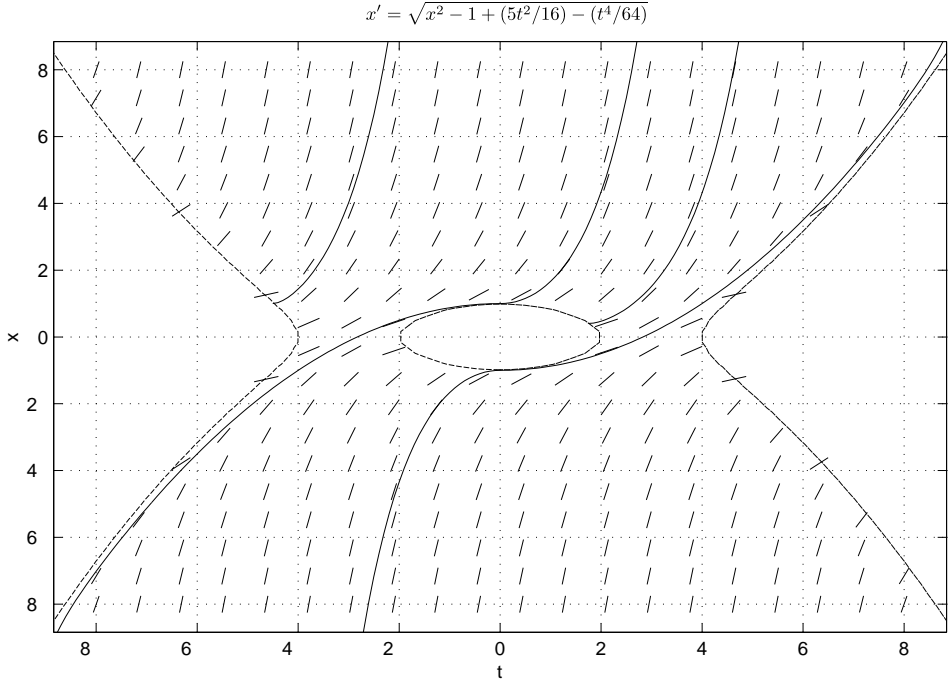,width=5in}}
    \caption{The $t$--$x$ plane, showing forbidden
regions.}\label{figTX} 
  \end{figure}
 A large-scale plot of the $t$--$x$ plane is shown in
figure~\ref{figTX}, and we see two large forbidden regions on the left
and right and a small one in the center. The inclined lines at a grid
of points are the slopes, obtained from \dXdT, of the solution curves
through each point.  The solution through $(t,x)=(0,1)$ is shown, and
it is easy to see that it gives the preferred superpotential $x_0(t) =
1- t^2/8$ only for $t<0$. This is related to the sign ambiguity of the
square root in \dXdT, and it is not a difficulty for us because we are
primarily concerned with the region $t<0$ which includes the full
range of the geometry containing two branes which was discussed in the
first part of this section.

Some other representative solutions are also plotted in
figure~\ref{figTX}. It is not proven, but it appears to be the case
that the only solutions which give a superpotential defined on the
full field space $-\infty<t<\infty$ are the curve through
$(t,x)=(0,1)$ and its mirror image through $(t,x)=(0,-1)$, which is
also shown in figure~\ref{figTX}.  Other solution curves reach the
boundary of the allowed region at a finite value of $t$ in one
direction, and one can see that $x'(t)$ vanishes but $x''(t)$ diverges
as one approaches the boundary.  By examining an approximate form of
\dXdT\ and \Weoms\ $\!,$ one can show that these curves approach the
boundary at a finite value of the coordinate $r$.  It then appears
that the solution curve reflects, and one must consider solutions of
\dXdT\ with the other sign of the square root.  The scale factor
$A(r)$ is smooth at the turning point.  This issue does not affect our
application, since the full brane geometry is contained in a region
without turning points.

Let's recall the logic of our construction. The potential $V(\phi)$
and left-hand brane tension $\lambda_1(\phi)$ are matched at a chosen
value $\phi=\phi_1$. We then choose the unique superpotential
$W(\phi)$ which satisfies $W(\phi_1) = \lambda_1(\phi_1)$ and agrees
in sign of slope with $\lambda_1'(\phi_1)$. Agreement in the magnitude
of the slope is guaranteed by (\ref{VWForm}) and
(\ref{VlambdaRel}). We then integrate the first order equations
\Weoms\ which gives the unique solution of the second order problem
\EOMS\ with the initial conditions $\phi(0)=\phi_1$,
$\phi'(0)=\lambda_1'(\phi_1)$, the latter from the jump condition
\Jump.  For consistency, it is useful to know that any other choice of
$W(\phi)$ leads to a different solution of \EOMS, one which does not
satisfy the jump conditions. This is quite clear from figure~\ref{figTX}, 
since the jump conditons. e.g. \WlambdaRel, are no longer
satisfied if we change solution chosen at the relevant fixed value
$t_1 =\sqrt{8/3}\, \phi_1$.

We have explored our suggested solution generating technique in only
one model.  Global issues associated with the turning points do not
spoil the applicability of our method, and the method is certainly
easy to use in the reverse mode where we start with a conveniently
chosen $W(\phi)$. We believe that this favorable situation is generic.

\section{Smooth solutions modeling branes}
\label{Smooth}

So far we have been considering solutions to an action that contains
explicit $\delta$-functions at the positions of the branes. One might
wonder to what extent this approach has already built in the answers
one wants to obtain. The purpose of this section is to present a
one-parameter family of purely 5d Lagrangians for gravity coupled to a
scalar, labelled by the parameter $\beta$, whose solutions are
generically smooth and asymptote to a specific $\delta$-brane solution
of the type considered so far. For generic $\beta$, the smoothed
branes appear as domain walls interpolating between various scalar
vacuua.  In the ``stiff'' limit ($\beta \rightarrow \infty$) the
second derivative of the scalar potential goes to infinity, so the
scalar becomes very heavy and can be integrated out.  The parameters
entering the scalar potential become the brane tensions and positions
associated with $\delta$-function terms in an action of the type
\SGRAV\ + \SBRANE\ after integrating out the scalar.

Several comments are in order. First, as mentioned before, we will not
be able to treat negative tension branes in this framework.  Second,
the solutions presented in this section do not have any fields living
on the brane, since the smooth solitons that in the stiff limit become
the branes do not have any zero modes. Both these obstacles can be
avoided by introducing ``by hand'' the $\delta$-functions in the action,
but this is precisely what we want to avoid with the smooth formalism.
In principle, the second limitation above could be overcome by
studying a more complicated smooth model which allows for non-trivial
zero modes on the brane.

Last but not least we should emphasize that even though we are
considering once more 5d gravity coupled to a scalar, this time the
scalar should not be thought of as the bulk scalar $\phi$ we studied
so far, which plays the role of a modulus for the fifth
dimension. Instead it is the scalar that the branes are made of!  In
order to avoid confusion we will call this auxiliary scalar $\xi$ and
reserve the symbol $\phi$ for the modulus scalar.  In the stiff limit,
where the soliton approaches the array of localized $\delta$-like
branes the fluctuations of $\xi$ are frozen out.  The bulk scalar
$\phi$ has to be introduced as a second scalar.  Interactions
localized on the brane, like the $\lambda(\phi)$ we introduced
earlier, can be mimicked by coupling the bulk scalar $\phi$ only to
derivatives of $\xi$.

We study a five-dimensional action of the form
\begin{equation}
S=\int_{M} d^4x \; dr \sqrt{|\mbox{det}g_{\mu \nu}}
\left [ -\frac{1}{4} R + \frac{1}{2} (\partial \xi)^2 - V(\xi)
\right ].
\end{equation}
We will work in the first order framework and hence take $V(\xi)$
to be given in terms of a ``superpotential''
as in (\ref{VWForm}) and study solutions to the first order equations
 (\ref{Weoms}). We will show that once we specify the potential
appropriately, the resulting solitonic solution describing an array
of branes with tension $\lambda_{\alpha}$ at positions $r_{\alpha}$ in the fifth
dimension is specified uniquely.

We are interested in the case where the scalar profile is given as
a solitonic domain wall configuration interpolating between various
vacua for the scalar field, e.g. written as
\begin{equation}
\label{profile}
\xi(r) =\sum_{\alpha} \frac{1}{\sqrt{\beta}} \;  \kappa_{\alpha} \;  \mbox{tanh} (\beta (r-r_{\alpha}) ),
\end{equation}
or a similar function that has the properties that
\begin{itemize}
\item in the ``stiff'' limit ($\beta \rightarrow \infty$) it reduces
to an array of step functions of height $\sim \frac{\kappa_{\alpha}}{\sqrt{\beta}}$,
and that
\item its first derivative is always negative and approaches a collection
of $\delta$-functions at position $r_{\alpha}$ of
strength $\sim \frac{\kappa_{\alpha}}{\sqrt{\beta}}$.
\end{itemize}
Note that latter property requires all $\kappa_{\alpha}$ to be positive,
ensuring that the function $\xi(r)$ is invertible. 
This solution in the stiff limit becomes an array of branes
of tension 
\begin{equation}
\lambda_{\alpha}=\frac{4}{3} \kappa_{\alpha}^2
\end{equation}
and only positive tensions appear.

Can we find a $W(\xi)$ in such a way that it allows a solution
of the form specified in (\ref{profile})? In order to do so, we just
rewrite the first order equation for the scalar flow in (\ref{Weoms})
as
\begin{eqnarray}
2 \xi'&=&\frac{\partial W(\xi)}{\partial \xi} = \frac{ \partial
W(\xi(r))}
{\partial r} \frac{ \partial r(\xi)} {\partial\xi} = \frac{W'}{\xi'} \\
\label{vofxi}
W(r) &=& 2 \int^r (\xi(r')')^2 \; dr'.
\end{eqnarray}
Using invertibility of $\xi(r)$ we can re-express $W(r)$ as $W(\xi)$
and hence obtain a potential $V(\xi)$ which leads to a solution of
the desired form. The one integration constant in $W$ corresponds to
an ``overall'' bulk cosmological constant. It should be chosen
in such a way that $A'(r)=-\frac{1}{3} W(r)$ is positive (negative)
to the left (right) of all branes. Since $A''$ is always negative,
it is always possible to choose the integration constant this way.
As we will see in the next section this property is enough to ensure that there
exists a 4-dimensional graviton.
Now we can turn the
philosophy around and say that once we have specified $V$ and hence
specified the action, or more precisely the bulk cosmological constant
and the cosmological constants between the various branes given in
terms of the value of $V(\xi)$ at its minima, the first order
equations then provide us with a solution of the form (\ref{profile})
for $\xi(r)$ together with the $A(r)$.
In the stiff limit this solution approaches an array of
sharply localized branes at
positions $r_{\alpha}$ and tensions $\lambda_{\alpha}$.

One should think of $V(\xi)$ as being obtained from
integrating out the microscopic physics. One then can ask again
whether there is some dynamical principle that determines the parameters
in $V$. Since we expressed $W$ as an integral over $(\xi')^2$ those
parameters are the $\kappa_{\alpha}$ and the $r_{\alpha}$.
Calculating the action integral of the solution as a function
of $\kappa_{\alpha}$ and $r_{\alpha}$
one finds
once more that it is always zero.
We remain with a serious fine-tuning problem:
the underlying theory has to be arranged in such a way, that for
given $\lambda_{\alpha}$ and $r_{\alpha}$ the potential has precisely the form
specified by (\ref{vofxi}). In the stiff limit all that remains
of $V$ are its values at the minima -- the inter-brane cosmological
constants\footnote{The normalization in (\ref{profile})
was chosen in such a way, that those inter-brane cosmological constants
remain finite in the stiff limit, $\frac{1}{L}$ jumps
by $\frac{8 \kappa^2}{9}$ when crossing a brane.} --
and the fine-tuning problem reduces to the standard fine-tuning of the
bulk cosmological constants against the brane tensions.

For example, in the case of a single brane we start with
\begin{equation}
\xi(r) = \frac{\kappa}{\sqrt{\beta}}\ \mbox{tanh} (\beta r) \,,
\end{equation}
leading to
\begin{equation}
\xi'(r) = \kappa \ \frac{\sqrt{\beta}}{\mbox{cosh}^2(\beta r)} \,,
\end{equation}
and hence
\begin{equation}
W=2 \kappa^2 \int (\xi')^2 \; dr= 2 \kappa^2 \left ( 
\mbox{tanh}(\beta r) -\frac{1}{3} \mbox{tanh}^3(\beta r) \right ) =
2 \kappa \sqrt{\beta} (\xi-\frac{\beta}{3 \kappa^2} \, \xi^3) \,. \label{WSolM}
\end{equation}
$A$ is simply obtained by integrating $W$.
In the multi-brane arrays the solution becomes slightly more complicated
due to the cross-terms in $(\xi')^2$ but it is still analytical.
One can show that in the stiff limit
all possible smoothings lead to the same brane array.

Before we end our discussion on smoothing of the singular solutions, let
us comment on how the coupling to the additional bulk scalar
looks in this framework.
In order
to mimic the localized interactions for the bulk scalar $\phi$ we
couple it to the derivatives of the auxiliary scalar $\xi$.
Basically, this means that we
couple a $\sigma$-model for the scalars to gravity, where the kinetic terms
of the auxiliary scalar $\xi$ depend on the bulk
scalar $\phi$. In the stiff limit
this once more will reduce to the solutions discussed in the previous sections.

Similar to (\ref{VWForm}) and (\ref{Weoms}) we can find a first order
formalism for the general action
  \eqn{NLSAction}{
   S=\int_{M} d^4x dr \sqrt{|\mbox{det}g_{\mu \nu}|}
    \left [ -\frac{1}{4} R + \frac{1}{2} G_{IJ} \partial^{\mu} \phi^J
    \partial_{\mu} \phi^I - V(\phi)
    \right ] \,,
  }
 where $G_{IJ}$ is a metric on the scalar target space.  Any solution to
  \eqn{WeomsTwo}{ 
   (\phi^I)' = {1 \over 2} \, G^{IJ} \, {\partial W(\phi) \over \partial\phi^J} \,,
    \qquad A' = -{1 \over 3} W(\phi) \,, 
  } 
 is also a
solution to the full second order equations provided $V$ is of the
special form 
  \eqn{VWFormTwo}{ 
   V(\phi) = {1 \over 8} \, G^{IJ} \, {\partial W(\phi) \over \partial \phi^I} \,
    {\partial W(\phi) \over \partial \phi^J} - {1 \over 3} W(\phi)^2\ .
  } 
 Choosing a two scalar model with $\phi^1=\phi$ and $\phi^2=\xi$ and
choosing $G_{12}=G_{21}=0$, $G_{11}=1$ and $G_{22}(\phi)$ to be an
arbitrary function of $\phi$ we should once more be able
(\ref{WeomsTwo}) to engineer a smooth model, this time limiting to
multi-brane-array in the presence of the bulk scalar $\phi$ with
localized interactions.

A count of parameters similar to the ones in section~\ref{Generate}
and~\ref{Nonzero} allows us to conclude that---at least locally---any
solution of the equations of motion following from \NLSAction\ which
preserves $3+1$-dimensional Poincar\'e invariance can be written as a
solution of \WeomsTwo\ for an appropriately chosen $W(\phi)$
satisfying \VWFormTwo.  Suppose there are $n$ scalars involved in the
action \NLSAction.  Each of them satisfies a second order equation of
motion.  The scale factor $A$ satisfies a first-order zero-energy
constraint analogous to the last line of \EOMS.  So there are $2n+1$
integration constants.  One of them can be absorbed into an additive
shift on $r$.  Now, \WeomsTwo\ leads to only $n+1$ integration
constants since the scalar equations are now first order.  But there
are also $n$ integration constants in \VWFormTwo\ regarded as a
partial differential equation for $W(\phi)$.  Again one integration
constant can be absorbed into an additive shift on $r$.  The point is
that either way we have the same number of integration constants, so
barring non-generic phenomena and global obstructions, the solution
spaces are the same.

This is quite an interesting result in view of the AdS/CFT
correspondence \cite{juanAdS,gkPol,witHolOne}.  One of the main
puzzles in the correspondence is how one might translate the
renormalization group (RG) equations, which are first order, into
supergravity equations, which are second order.  In \cite{fgpw1} first
order equations were extracted from the conditions for unbroken
supersymmetry.  These equations are suggestive of an RG flow
based on the gradient of a c-function.  The c-function is $W(\phi)$,
and its relation to the conformal anomaly arises because of the
equation $A' = -{1 \over 3} W$: in regions where the scalars are
nearly constant and the geometry is nearly $AdS_5$, an application of
the analysis of \cite{hs} shows that the Weyl anomaly coefficients in
the conformal field theory are proportional to the third power of the
radius of $AdS_5$, or equivalently to $|W|^{-3}$.  (Thus in a sense it would
be more appropriate to speak of $|W|^{-3}$ as the c-function.)

In a non-supersymmetric ``flow,'' the c-function can still be defined
\cite{GPPZ,fgpw1} as $-3 A'$, and it is possible to demonstrate $A'' \leq
0$ using only the weakest of positive energy conditions \cite{fgpw1}.  But
then the spirit of RG is lost: one wants to have a notion of a first order
flow through the space of possible theories labelled by different values of
parameters, and whatever c-function one constructs should be defined in
terms of those parameters.  The construction of $W$ indicated in
\VWFormTwo\ seems to realize this idea explicitly.

However there are some caveats.  First, $W$ depends on $n$ integration
constants, where as before $n$ is the number of scalars.  It seems
reasonable that these integration constants can be interpreted as
specifying the state of the dual field theory, which does not change under
RG---only the Hamiltonian evolves.  Second, the same phenomena of forbidden
regions and turning points that we discussed in section~\ref{Numerics}
occur also in the case of several scalars.  A forbidden region is a region
of $(W,\phi)$ space where $V(\phi) + {1 \over 3} W^2$ is negative.  Barring
singular behavior in $G_{IJ}$, one finds that the gradient of $W$ vanishes
at the border of these regions, so no flow can cross over.  Rather, flows
reflect from the border and the subsequent flow is controlled by a
different branch of $W$.  Because of the multi-valued nature of $W$, we do
not regard \WeomsTwo\ as a wholly satisfactory starting point for the
transcription of supergravity equations into RG equations.  However it is
perhaps a step in the right direction.

\section{Fluctuations around the solution}
\label{Fluctuate}

Finally we examine the equations governing fluctuations of the metric
and scalar around the classical background solutions of the equations
of motion of the action \Sfive.  Our methods are somewhat
different from those in the literature.  We choose an axial-type
gauge, and the resulting form of the four-dimensional graviton is
particularly simple.  Transverse traceless modes in general obey the
equation of a massless scalar in the curved background, and by
recasting this as the Schr\"odinger equation for a supersymmetric
quantum mechanics problem, we argue that there are no space-like modes
threatening stability.

We impose the ``axial gauge'' constraint, so named for its resemblance
to $A_3 = 0$ in electrodynamics:
\eqn{AxialGauge}{ h_{\mu 5} = 0 \,,} where $\mu = 0,1,2,3,5$.
We can then write the total metric in the form
\eqn{PerturbedMetric} {ds^2 = e^{2 A(r)} (\eta_{ij} + h_{ij}) \, dx^i dx^j - dr^2
\,,}
where we extracted a factor $e^{2A}$ from the fluctuation term to
simplify future equations.  Axial gauge is not a total gauge fix, as
diffeomorphisms generated by a vector field $\epsilon_i = e^{2A(r)} \,
\omega_i(x^j)$, $\epsilon_5 = 0$ preserve the condition \AxialGauge\
while transforming the fluctuations $h_{ij}$ as \eqn{FourDGauge}{
h_{ij}(x^k, r) \rightarrow h_{ij}(x^k, r) + \partial_i \,
\omega_j(x^k) +\partial_j \, \omega_i(x^k) \,.  }  Note the
resemblance to four-dimensional diffeomorphisms. \footnote{There is
a more general residual gauge invariance involving  a non-vanishing
$\epsilon_5(x^k)$. See  \cite{GT}.}

The Ricci tensor can be computed from the metric
\PerturbedMetric.  To zeroth order in the fluctuations we
continue to have \ricciFive, while to first order we calculate
(using Maple):
\eqn{PerturbedRicci}{\eqalign{ R_{ij}^{\: (1)} &= e^{2 A} \left(
\frac{1}{2} \partial_r^2 + 2 A' \partial_r + A'' + 4 A'^2 \right)
h_{ij} + {1 \over 2} \, \eta_{ij} \, e^{2A} A' \, \partial_r
(\eta^{kl} h_{kl}) - {1 \over 2} \square h_{ij}  \cr &\quad - \frac{1}{2} \, \eta^{kl} \left(
 \partial_i \partial_j h_{kl} -
 \partial_i \partial_k h_{jl} - \partial_j \partial_k h_{il} 
\right) \,, \cr R_{55}^{\: (1)} &= - {1 \over 2} (\partial_r^2 + 2 A'
\partial_r) \, \eta^{kl} h_{kl} \,, \quad \quad R_{j5}^{\: (1)} = {1
\over 2}\, \eta^{kl}\, \partial_r\, (\partial_k h_{jl} - \partial_j
h_{kl}) \,,  }}
where $\square = \eta^{ij} \partial_i \partial_j$ is the flat four-dimensional Laplacian.  
Einstein's equations in Ricci form require that $R_{\mu \nu} =
\bar{T}_{\mu \nu} \equiv T_{\mu \nu} - {1 \over 3} \, g_{\mu \nu} \,
T^{\, \alpha}_{\; \; \, \alpha}$, and we find
  \eqn{PerturbedEMTensor}{\eqalign{\bar{T}_{ij}^{\: (1)} &= -{2 \over 3}\,
  e^{2A} \Bigg[ \left(2 {\partial V(\phi) \over \partial \phi} + \sum_{\alpha} {\partial
  \lambda_{\alpha}(\phi) \over \partial \phi} \delta(r - r_{\alpha})
  \right) \tilde{\phi}\, \eta_{ij}  \cr & \quad + \left(2 V(\phi) + \sum_{\alpha}
  \lambda_{\alpha}(\phi) \delta(r - r_{\alpha}) \right) h_{ij} \Bigg]
  \cr \bar{T}_{55}^{\: (1)} &= 4 \phi' \tilde{\phi}' + {4 \over 3} \left(
  {\partial V(\phi) \over \partial \phi} + 2 \sum_{\alpha} {\partial \lambda_{\alpha} \over \partial
  \phi} \delta(r - r_{\alpha}) \right) \tilde{\phi} \,, \quad \quad
  \bar{T}_{j5}^{\: (1)} = 2 \phi' \, \partial_j \tilde\phi \,. }} 
Additionally, the equation of motion for the scalar fluctuation
$\tilde{\phi}$ is
  \eqn{PerturbedScalar}{ e^{-2A} \, \square \, \tilde{\phi} -
  \tilde{\phi}'' - 4 A' \tilde{\phi}' + \left( {\partial^2 V(\phi) \over
\partial \phi^2} + \sum_{\alpha} {\partial^2 \lambda(\phi) \over
\partial^2 \phi} \delta(r - r_{\alpha}) \right) \tilde{\phi} = {1
  \over 2} \phi' \eta^{ij} h_{ij}' \,.}
The equation $R_{ij}^{\: (1)} = \bar{T}_{ij}^{\: (1)}$ 
further simplifies as a consequence of the zeroth-order equation of
motion (\ref{EOMS}) : \eqn{Cancel}{ A'' + 4 A'^2 = - {4 \over 3}
V(\phi) - {2 \over 3} \sum_{\alpha} \lambda_{\alpha}(\phi) \delta(r -
r_{\alpha}) \,, } to \eqn{RicciOne}{\eqalign{ e^{2 A} \left(
\frac{1}{2} \partial_r^2 + 2 A' \partial_r \right) h_{ij} - {1 \over
2} \square &h_{ij} \cr + {1 \over 2} \, \eta_{ij} \, e^{2A} A' \,
\partial_r (&\eta^{kl} h_{kl}) - \frac{1}{2} \, \eta^{kl} \left(
\partial_i \partial_j h_{kl} - \partial_i \partial_k h_{jl} -
\partial_j \partial_k h_{il} \right) = \cr &-{2 \over 3}\, e^{2A}
\left(2 {\partial V(\phi) \over \partial \phi} + \sum_{\alpha} {\partial \lambda_{\alpha}(\phi)
\over \partial \phi} \delta(r - r_{\alpha}) \right) \tilde{\phi}\,
\eta_{ij} \,.}}
Let us now consider the transverse traceless components of $h_{ij}$,
defined by the non-local projection \cite{Anselmi}:
\eqn{Projection}{\eqalign{ \overline{h}_{ij} = 
\left( {1 \over 2} ( \pi_{ik} \, \pi_{jl} + \pi_{il} \, \pi_{jk} )
- {1 \over 3} \pi_{ij} \, \pi_{kl} \right) \, h^{kl} 
= h_{ij} + \ldots \,,
}}
where $\pi_{ij} \equiv (\eta_{ij} - \partial_i \partial_j / \square)$
and $\ldots$ indicates nonlocal terms.
The $\overline{h}_{ij}$ satisfy \eqn{TT}{ \partial^j
\overline{h}_{ij} = \eta^{ij} \, \overline{h}_{ij} = 0 \,.  }  We
emphasize that $\TT$ applies only to the components defined in
\Projection\, and is not a gauge choice; it would be incompatible with
\AxialGauge\ and the residual gauge freedom \FourDGauge.

For the $\overline{h}_{ij}$, \RicciOne\ simplifies enormously.
The transverse traceless projection removes the right-hand side,
and we are left with
\eqn{TTEqn}{ \left( \partial_r^2 + 4 A' \partial_r - e^{-2A}
\square \right) \overline{h}_{ij} = 0 \,.  }
Notice that all $\delta$-function jumps have canceled out;
this is nothing but the equation of motion for a free massless scalar
in our curved background.  
In an $AdS_5$ black hole background, the spin-2 components of the
graviton were also found to obey a free scalar wave equation
\cite{Neil,BMT}.

We expect one solution of our equations to be the four-dimensional
graviton.  Since it is massless in the four-dimensional sense, it must
obey $\square \overline{h}_{ij} = 0$.  We can easily see that such a
solution to \TTEqn\ is simply the $r$-independent plane wave
\eqn{FourDGraviton}{ \overline{h}_{ij} = C_{ij} \, e^{ipx} \,, } where
$p^2 = 0$ and $C_{ij}$ is a constant.  Thus in this presentation the
phenomenological graviton has a very simple form.

As we will argue below, the norm of
metric fluctuations is \eqn{rConverge}{ || \, \overline{h}\, ||^2 =
\int dr \,
e^{2A(r)}\, \overline{h}_{ij} \, \overline{h}^{ij} \,, } 
where indices are raised with $\eta^{ij}$.
We see that the graviton mode \FourDGraviton\ is
normalizable because the $r$-direction is effectively compactified in
these models.  The $S^1/Z^2$ geometries are manifestly compact.  For
arrays of positive-tension branes only, the range of $r$ is $-~\infty~<~r~<~\infty$, but the norm converges if we restrict to cases where
\eqn{AsymADS}{\eqalign{ A' \rightarrow 1/L_- >0 \quad &{\rm as} \quad r
\rightarrow - \infty \,, \cr A' \rightarrow - 1/L_+ <0 \quad &{\rm as}
\quad r \rightarrow \infty \,,}}
which are asymptotically anti-de Sitter geometries.  In all such
models, which include the smooth configurations of section
\ref{Smooth}, there is a naturally massless four-dimensional graviton
as described above.

Having identified the four-dimensional graviton, we next turn to the
question of stability.  If the equations of motion were to admit
fluctuations with a space-like momentum, it would be evident that the
zeroth-order solution --- our classical background --- is not stable.
For the transverse traceless components, we can cast the expression
\TTEqn\ in the form of a supersymmetric quantum mechanics problem,
where $p^2$ plays the role of the energy, and thus argue that $p^2
\geq 0$.

To accomplish this, we first need to eliminate the factor $e^{2A}$
multiplying the momentum.  We can do this by changing variables
to coordinates in which the background is conformally flat:
\eqn{ConformallyFlat}{ds^2 = e^{2A(z)} \left( (\eta_{ij} + h_{ij}) \, dx^i dx^j - dz^2 \right) \,.}
Now \TTEqn\ takes the form
\eqn{ConFlatEqn}{(-\partial_z^2 - 3 \, A'(z)\, \partial_z + \square) \, \overline{h}_{ij} =0\,. }
In terms of
$H_{ij}(z) =e^{-ipx} \, e^{3A/2} \, h_{ij}$, this becomes
\eqn{SUSYQM}{
\left( - \partial_z^2 +  {9 \over 4} A'(z)^2 + {3 \over 2} A''(z) \right) H_{ij}(z) = p^2 H_{ij}(z) \,. }
This differential operator has the same form as a Hamiltonian in quantum
mechanics, with a potential $V(z) = \tf{9}{4} A'(z)^2 + \tf{3}{2} A''(z)$
and $p^2$ as the energy eigenvalue.  One can easily check that
it factorizes
\eqn{QQbar}{\left( (\partial_z + {3 \over 2} A'(z))  (-\partial_z + {3 \over 2} A'(z)) \right) H_{ij}(z) = p^2 H_{ij}(z) \,.}
In flat space, these terms are one another's adjoint, and \QQbar\
can be regarded as a factorization of the Hamiltonian into 
$\overline{Q}Q$.  This is supersymmetric quantum mechanics, and the
transformed graviton wave-function is the supersymmetric ground state.
However, to complete the argument we must show that a flat-space norm
is correct for $H_{ij}(z)$ in our curved background.

In Lorentzian signature field theory, the norm of fluctuations is 
determined by the requirement that formally conserved quantities such as the
contraction $T^{\mu\nu} K_\nu$ of the stress tensor and a Killing vector of
the background have convergent integral
\eqn{ConvInt} { \int dz \, d^3 x \, \sqrt{g} \, T^{0\nu} \, K_{\nu} \,, }
over a constant time 4-surface and vanishing flux through its boundary
3-surface. Stress tensors for metric fluctuations are complicated, but in
this linearized situation the stress tensor must be covariantly conserved
for all solutions of the equation of motion \TTEqn\ or \ConFlatEqn\ -
\SUSYQM\ . Thus for the
Killing vector ($K^0=0$, $K^i=$constant, $K^5=0$) of spatial translations
parallel to the domain wall, one can take the form
\eqn{ConsStrEn}{T^0_i = e^{-2A} \, \partial_0 \overline{h}_{kl} \, \partial_i
\overline{h}^{kl} \,, } 
obtained by specializing the obvious covariant expression for $T^0_i$ to our
description of the background. (The index $i$ takes values 1,2,3 in 
\ConsStrEn\,
while $k$, $l$ are raised with $\eta^{kl}$.) The requirement of a convergent
integral for the spatial momentum carried by the fluctuation then constrains
the radial eigenfunctions $H_{ij}(z)$ to satisfy\footnote{
We thank the authors of \cite{CsakiErlichHollowood} for pointing out that
our initial discussion of the norm was incorrect. The correct norm appears
in \cite{CsakiErlichHollowood} and elsewhere; see, for example 
\cite{brandhuber,CohenKaplan}.}
\eqn{RightNorm}{
\int dz \,
H_{ij} \, H^{ij} \, = \mbox{ finite}.} 
which is the usual Schr\"odinger norm for \SUSYQM\ 
(and equivalent to \rConverge\ when
rephrased in terms of $\overline{h}_{ij}$ and the radial coordinate $r$). 
Supersymmetric quantum
mechanics thus ensures that there are no normalizable modes with $p^2<0$.
Thus we can state that there are no transverse traceless modes with
space-like momentum that might destabilize the backgound solution.

Before concluding this section, we briefly remark on the 
non-transverse traceless components of the metric fluctuation,
which are coupled to the scalar by the equations
\PerturbedRicci, \PerturbedEMTensor, and \PerturbedScalar.
These coupled equations are not easy to solve, and we have not attempted
to rule out tachyonic modes of these fluctuations here.

However, it seems likely that the Boucher non-supersymmetric
positive-energy theorem \cite{Boucher, Townsend} can be extended to
include actions such as ours with potentials localized on
hypersurfaces, in which case stability would be guaranteed for our
solutions, by virtue of their satisfying the first-order equations.

\section*{Note Added}

As this manuscript was nearing completion, several papers appeared
\cite{KimSquared,Hatanaka,BehrndtCvetic,Skenderis} which overlap somewhat
with our results.  For instance, \LagSimp\ was also derived in
\cite{Skenderis}, and the $dS_4$ solution in \deSitterSol\ was also
obtained in \cite{KimSquared}.  In \cite{BehrndtCvetic}, solutions similar
to the single domain wall of section~\ref{Smooth} were shown to emerge from
a $U(1)$ gauged supergravity theory.

The coupled equations relating scalar and non-transverse metric
fluctuations have recently been studied in \cite{dewolfe-freed}. The
equations can again be reduced to the form of supersymmetric quantum
mechanics, and consequently there are no normalizable spacelike modes.
Thus our backgrounds have been shown to be entirely free from
tachyonic fluctuations.

\section*{Acknowledgements}

We would like to thank D.~Gross, S.~Kachru, R.~Myers, V.~Periwal, and M.~Perry
for useful discussions.  The research of D.Z.F.\ and was supported in
part by the NSF under grant number PHY-97-22072.  The research of
O.D.\ and A.K.\ was supported by the U.S.\ Department of Energy under
contract \#DE-FC02-94ER40818.  The research of S.S.G.\ was supported
by the Harvard Society of Fellows, and also in part by the NSF under
grant number PHY-98-02709, and by DOE grant DE-FGO2-91ER40654.
D.Z.F., S.S.G., and A.K.\ thank the Aspen Center for Physics for
hospitality.

\bibliography{wise}
\bibliographystyle{ssg}

\end{document}